\documentclass[aps,twocolumn,superscriptaddress,longbibliography]{revtex4-1}
\setcitestyle{super}

\AtBeginDocument{}

\usepackage{tikz}
\usepackage{lipsum}
\usepackage{graphicx}
\usepackage[export]{adjustbox}
\usepackage{amsmath,amssymb,wasysym,amsthm,bm,dsfont}
\usepackage{braket}
\usepackage{mathtools}
\usepackage{mathrsfs}
\usepackage{verbatim}
\usepackage{makecell}
\usepackage{cases}
\usepackage{hyperref}
\usepackage{diagbox}

\newcommand{\im}{\mathbf{i}}

\newcommand{\abs}[1]{\left| #1\right|}%%adjustable-height norm shortcut
\newcommand{\Rev}[1]{{#1}}
\newcommand{\red}[1]{{#1}}

\newcommand{\fig}[1]{Fig.~\ref{fig:#1}}
\newcommand{\eq}[1]{Eq.~\ref{eq:#1}}
\newcommand{\sect}[1]{Section.~\ref{#1}}
\newcommand{\apdx}[1]{Appendix.~\ref{#1}}

\newlength{\wdth}

\begin{document}
\title{A symmetry-protected topological optical lattice clock}
%\title{Topologically robust sensing via engineering the Su-Schrieffer-Heeger model in a one dimensional optical lattice clock}
\author{Tianrui Xu}
\affiliation{JILA, NIST and Department of Physics, University of Colorado, Boulder, Colorado 80309, USA}
\affiliation{Center for Theory of Quantum Matter, University of Colorado, Boulder, Colorado 80309, USA}
\affiliation{Institut Quantique and D\'epartement de Physique, Universit\'e de Sherbrooke, Sherbrooke, Qu\'ebec J1K 2R1, Canada}
\author{Anjun Chu}
\affiliation{JILA, NIST and Department of Physics, University of Colorado, Boulder, Colorado 80309, USA}
\affiliation{Center for Theory of Quantum Matter, University of Colorado, Boulder, Colorado 80309, USA}
\author{Kyungtae Kim}
\affiliation{JILA, NIST and Department of Physics, University of Colorado, Boulder, Colorado 80309, USA}
\author{James K. Thompson}
\affiliation{JILA, NIST and Department of Physics, University of Colorado, Boulder, Colorado 80309, USA}
\author{Jun Ye}
\affiliation{JILA, NIST and Department of Physics, University of Colorado, Boulder, Colorado 80309, USA}
\author{Tilman Esslinger}
\affiliation{Institute for Quantum Electronics \& Quantum Center, ETH Zurich, 8093 Zurich, Switzerland}
\author{Ana Maria Rey}
\affiliation{JILA, NIST and Department of Physics, University of Colorado, Boulder, Colorado 80309, USA}
\affiliation{Center for Theory of Quantum Matter, University of Colorado, Boulder, Colorado 80309, USA}

\begin{abstract}
We theoretically propose a tunable implementation of symmetry-protected topological phases in a synthetic superlattice, taking advantage of the long coherence time and exquisite spectral resolutions offered by gravity-tilted optical lattice clocks. We describe a protocol similar to Rabi spectroscopy that can be used to probe the distinct topological properties of our system. We then demonstrate how the sensitivity of clocks and interferometers can be improved by the protection to unwanted experimental imperfections offered by the underlying topological robustness. The proposed implementation opens a path to exploit the unique opportunities offered by symmetry-protected topological phases in state-of-the-art quantum sensors.
\end{abstract}

\maketitle

\section{Introduction}
Recent years have witnessed rapid and exciting new developments of optical lattice clocks (OLCs) with excellent quantum coherence and exquisite spectral resolutions\cite{nicholson15natcomm,clock15rmp,campbell17sci,mcgrew18nat,oelker19nat,bothwell2022,zheng22nat,aeppli22tiltSr,Beloy2021,Aeppli2024}. Such platforms are ideal for quantum sensing, and have recently reached clock measurement precision at $7.6\times10^{-21}$ and near minute-long atomic coherence\cite{bothwell2022,Aeppli2024}, making it possible to precisely measure quantities of small magnitudes, such as the gravitational redshift across a millimeter- to centimeter-length scale\cite{bothwell2022,zheng22nat}.

In parallel, over the past decades, the rapid development of quantum simulation with cold atomic systems has enabled experimental investigations of topological properties of quantum matter.\cite{cooper19rmp,citro23natrev} In particular, significant progress has been made to experimentally realize a class of topological states of matter, referred to as ``symmetry-protected topological (SPT) phases.'' This type of quantum matter has the property of being insulating in the bulk (i.e., having a gapped dispersion) while conducting at the boundary (i.e., having a gapless dispersion) as long as a certain global symmetry is preserved. Namely, the SPT phases are robust to perturbations that respect said global symmetry and do not close the bulk gap. A prototypical model of a SPT system is the celebrated Su-Schrieffer-Heeger (SSH) model\cite{ssh79,ssh80}. This model, and its closely related models such as the Rice-Mele (RM) model\cite{ricemele}, have been realized in a variety of settings including  superlattices\cite{atala13natphys,lohse16natphys,nakajima16natphys,walter23natphys}, momentum-space lattices\cite{meier16natcomm,meier18sci,yuan23prr}, Rydberg atoms\cite{rydbergssh19} and multilevel systems\cite{kanungo22natcomm}. Many interesting properties and dynamics of the SSH/RM model have been observed in these experiments, such as soliton/edge state dynamics\cite{meier16natcomm,leder16natcomm,rydbergssh19,katz2024observing}, Zak phase measurements\cite{atala13natphys,meier18sci,xie19npj}, Thouless pumping\cite{lohse16natphys,nakajima16natphys, lu16prl,walter23natphys}, edge-to-edge transport\cite{yuan23prr} and topological quantum walks\cite{meier18sci}. 

At the moment, however, quantum simulations of topological quantum matter appear to be independent from the field of quantum sensing. While the realization of topological phases is exciting in its own right, what is even more appealing is the potential use of topological robustness to remove \Rev{the} vulnerability of \Rev{state-of-the-art} sensors to unwanted noise while keeping their sensitivity to the desired signal. \Rev{\red{In fact, state-of-the-art optical lattice clocks are limited by the Dick (sampling) noise of an optical local oscillator, and a protocol that extends the coherent evolution time would be highly valuable. Moreover, even in synchronous comparison between two clocks, which are in principle immune to the Dick noise, at the current $10^5$ atoms, the differential measurements are still not yet limited by the standard quantum limit (SQL)} -- the fundamental bound in sensitivity achievable with uncorrelated atoms. Therefore, an urgent current need is to find noise-suppressing protocols to allow current clocks to operate at the SQL, before implementing strategies to further improving OLCs via quantum entanglement.}

\begin{figure*}
    \centering
    \includegraphics[width=\linewidth]{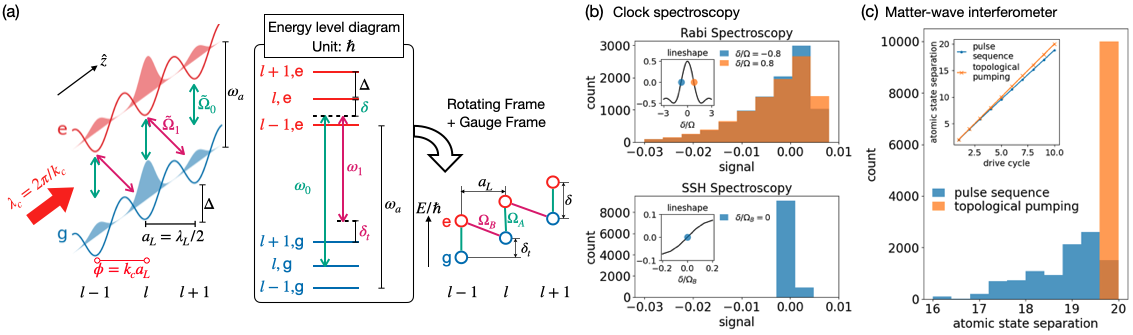}
    \setlength{\abovecaptionskip}{-7pt}
    \setlength{\belowcaptionskip}{-7pt}
    \caption{A symmetry-protected topological OLC. (a): Simulating the SSH/RM models in a tilted 1D OLC: by using two laser beams to drive both carrier and sideband transitions, a 1D Wannier-Stark ladder can be transformed into the SSH/RM model that we use for quantum simulation and quantum sensing protocols discussed in the rest of the manuscript. Left panel: a schematic diagram of the experimental realization of $\hat H_\text{t-RM}$ in a tilted 1D OLC. We use a two-tone clock laser with wavevector $k_c$ in a tilted 1D OLC with lattice constant $a_L$. When the clock laser is incommensurate with the optical lattice, atoms on neighboring lattice sites feel a spin-orbit-couping phase $\phi=k_ca_L$. We show the Wannier-Stark wavefunctions as shaded red and blue areas. Center panel: the energy diagram relevant to generate the SSH\Rev{/RM} model, as indicated in the right panel. \Rev{(b) and (c): Reducing undesirable noise via topologically robust protocols. Both panels include experimentally-relevant, shot-to-shot noise (see also Sections IV and V): Here we examine $100$ realizations of global laser amplitude noise and $100$ realizations local laser amplitude noise, assuming $0.1\%$ global noise and $4\%$ local noise, respectively. (b): Clock spectroscopy via Rabi (upper panel) and SSH (lower panel) protocols. The main panels show histograms of the signals at two opposite detunings as indicated in the legends and illustrated with the colored symbols in the insets. The insets in the upper and lower panels show the lineshapes of the corresponding protocols. (c): Spatial separation of the  $e$ and $g$ atomic states in an OLC interferometer via a sequence of  pulses (blue) and via an adiabatic, topological Thouless pumping scheme (orange). The main panels show the histograms of the atomic state separation after $10$ drive cycles. The insets show the averaged spatial separation of the $e$ and $g$ states at different number of cycles.}}
    \label{fig:sys}
\end{figure*}

\Rev{Here we propose a possible path forward via topological states that are relatively easy to implement in state-of-the-art non-interacting OLCs\cite{Aeppli2024}, which can provide metrological benefits from their underlying topological robustness. More specifically, we discuss} a protocol to create the simplest \Rev{SPT} model, the SSH model, in a tilted one-dimensional (1D) OLC (see \fig{sys}-(a)). The tilt can come from the gravitational acceleration or other types of acceleration, parallel to the 1D OLC axis\cite{bothwell2022,aeppli22tiltSr,Aeppli2024,zheng22nat}. 

\Rev{In our proposed setting, atoms are allowed to tunnel in a highly-controlled manner without energy costs, leading to spatially-delocalized atomic wavefunctions across the lattice. In the topologically non-trivial phase, specific observables of interest become insensitive to variations of the Hamiltonian parameters that describe the atom dynamics. Such robustness has already been observed in previous experiments\cite{meier18sci}. It originates from the Berry phase\cite{zakprlberry} (or the geometric phase) that atoms accumulate when moving through the lattice. This is similar to the quantized and non-zero Hall conductivity in certain solid state systems, where the Hall conductivity is robust against noise and fluctuations in material parameters.} 

\Rev{In our system, such topological robustness protects coherences in highly mobile atoms, mitigating the so-far detrimental decoherence from the spin-orbit coupling in tunneling-dominated regimes\cite{kolkowitz17nat}. It allows to perform clock spectroscopy in a way that is not only robust to local noise -- which averages out as atoms delocalize, but also more resilient to global noise sources compared to standard frozen atoms. \fig{sys}-(b) illustrates how the topological robustness of our system can help suppress the OLCs' sensitivity to Rabi frequency amplitude noise. The upper panel shows how standard Rabi spectroscopy is impacted by the presence of experimentally-relevant amplitude noise that we discuss in more detail in \sect{sshspec}. The lower panel shows the robustness to such noise enabled by the ``SSH Spectroscopy'' protocol, also discussed in \sect{sshspec}.

Furthermore, we show that the capability of atoms to quickly delocalize, even in the presence of a gravitational tilt, enables our protocol to be useful not only for optical phase estimations, but also as a lattice-based matter-wave interferometer (MWI)\cite{panda24nat}. Such capability is made possible by being able to achieve large spatial separation of wave packets. Naively speaking, while a sequence of $\pi$-pulses\cite{pelle13pra,chu21prl} can be used to spatially separate the wave packets, as illustrated in \fig{sys}-(c), pulse imperfections from Rabi frequency amplitude noise can both generate a large uncertainty and reduce the value of the achievable separation. These issues can be mitigated using adiabatic topological pumping schemes (detailed in \sect{topomwi}). As schematically illustrated in this panel, the spatial separation generated via adiabatic topological pumping methods can reach larger values with reduced uncertainty.

We emphasize that, given the non-interacting conditions in consideration, our protocols do not generate entanglement, and therefore do not reduce quantum noise. Instead, our protocols help reduce the \emph{classical statistical noise} caused by relevant experimental imperfections, opening a path for state-of-the-art clocks operating at the desired SQL sensitivity.

More specifically, we study the performance of our quantum sensing protocols in the presence of experimentally-relevant static noise, including laser amplitude noise and noise arising from atomic motion and thermal effects\cite{yan25,blatt09pra,kim23prl}. We use realistic parameters and approximations in our analysis.}

\Rev{The rest of this manuscript is organized as follows. First, we discuss a protocol capable} to engineer the SSH model with tunable model parameters in \Rev{tilted} OLCs. \Rev{Then, }we discuss how to \Rev{use clock spectroscopy to }probe and characterize the different topological phases of the SSH model. Finally, we discuss how topology protects the system's sensitivity against unwanted noise in quantum sensing protocols of \Rev{(1) an optical frequency and (2) a constant uniform acceleration}.

Our work can be readily implemented in current experimental platforms\cite{aeppli22tiltSr,Aeppli2024} and paves ways to study topological phases of quantum matter in OLCs. More importantly, our work opens up a path to use topological protection to improve clock operations.

\section{The Su-Schrieffer-Heeger model in a tilted OLC}

In this section, we discuss our proposal to simulate the SSH/RM model, \Rev{a} prototypical model featuring \Rev{SPT} phases, in a 1D tilted OLC. We then briefly review the key concepts of this model relevant for the rest of this manuscript.

We consider a tilted, 1D OLC with lattice spacing $a_L$ and nearest-neighbor tunneling frequency $J$, loaded with a dilute array of atoms in such a way the system can be considered as non-interacting, as shown in \fig{sys}-(a). The atoms have mass $M_a$, and experience a uniform acceleration $g_{acc}$ generated by a linear potential across the lattice. The eigenstates of this system are the Wannier-Stark (WS) states, $\ket{l}$\cite{lemonde05pra,aeppli22tiltSr}, centered at lattice site $l$ and with eigenenergy $E_l=\hbar\Delta l$, where $\hbar\Delta\equiv M_ag_{acc}a_L$ is the energy difference between atoms on adjacent lattice sites. We consider a parameter regime where the lattice depth is shallow enough that $J$ is comparable to $\Delta$, thus $\ket{l}$ is delocalized across several lattice sites. We show the WS states as shaded areas in the left panel of \fig{sys}-(a). The clock states correspond to two electronic levels, $e$ and $g$, with an optical transition frequency $\omega_a$. The single-particle Hamiltonian of our system thus reads:
\begin{align}
    \hat H^\text{1p}_\text{lab}=\hbar\Delta\sum_{l,\alpha=g,e}l\hat{\tilde c}^\dagger_{l\alpha}\hat{\tilde c}_{l\alpha}+\sum_l\hbar\omega_a\hat{\tilde c}^\dagger_{le}\hat{\tilde c}_{le},
\end{align} where $\hat{\tilde c}_{l\alpha}$ annihilates a fermionic particle in  WS state $\ket{l}$  and electronic level $\alpha =g,e$. \Rev{It is worth noting that the SSH/RM model is a single-particle model, therefore independent of quantum statistics. However, in reality, Fermi statistics makes it easier to implement our protocol both by reducing the role played by interactions and for the initial state preparation.}

We use a clock laser with wavelength $\lambda_c$, propagating along the lattice direction, to drive the ultranarrow clock transition $g-e$. When $\lambda_c/a_L$ is not an integer value, the clock laser imprints a differential phase $\phi=k_ca_L$ across adjacent lattice sites which generates the spin-orbit coupling (SOC)\cite{wall16prl,livi16prl,kolkowitz17nat,zhang19jpcs}, allowing the clock laser to drive $g-e$ transitions between different sites. We propose to use two different clock laser tones to implement the SSH/RM model. The first tone with angular frequency $\omega_0$, detuned from the atomic transition by $\delta=\omega_0-\omega_a$, drives the on-site (``carrier'', $\ket{l,g}\to\ket{l,e}$) transition (see \fig{sys}-(a) and \apdx{fullhami} for details):
\begin{align}
    \hat H^{\tilde\Omega_0}_\text{lab}&=\frac{\hbar \tilde\Omega_0}{2}e^{-\im\omega_0t}\sum_l\hat{\tilde c}^\dagger_{le}\hat{\tilde c}_{lg}+\text{h.c.},
\end{align}
where $\tilde\Omega_0= \Omega^c \langle l| e^{ik_c\hat x}|l\rangle\simeq\Omega^cI_{0}\mathcal{J}_0(\tilde J)$ with $\Omega^{c}$ the bare Rabi frequency of the carrier drive and $I_0=\int dxe^{\im kx}w^2_0(x)$  the on-site overlapping integral of the localized ground-band Wannier function $w_0(x)$, and  $\mathcal J_n(\tilde J)$ the $n$-th Bessel function of the first kind with $\tilde J=4J\abs{\sin(\phi/2)}/\Delta$. \Rev{The second tone with angular frequency $\omega_1$ is used to  drive a ``red sideband'' transition, $\ket{l,g}\to\ket{l-1,e}$, with two-photon detuning $\delta_t=\omega_1-\omega_0+\Delta$. The Hamiltonian thus reads }% This tone has a two-photon detuning $\omega_1-\omega_0= \delta_t-\Delta$ and the Hamiltonian reads 
(see \fig{sys}-(a) and \apdx{fullhami} for more details):
\begin{align}
    \hat H^{\tilde\Omega_1}_\text{lab}&=\frac{\hbar \tilde\Omega_1}{2}e^{-\im\omega_1t}\sum_l\hat{\tilde c}^\dagger_{le}\hat{\tilde c}_{l+1g}+\text{h.c.},
\end{align}
where $\tilde\Omega_1= \Omega^s \langle l-1| e^{ik_c\hat{x}}|l\rangle\simeq\im\Omega^sI_{0}\mathcal J_{-1}(\tilde J)$ and $\Omega^{s}$ is the bare Rabi frequency of the sideband drive. In the above equations, we removed the SOC phase via a gauge transformation: \Rev{$\hat{\tilde c}^\dagger_{le}\to e^{-\im l\phi/2}\hat{\tilde c}^\dagger_{le}$, $\hat{\tilde c}^\dagger_{lg}\to e^{\im l\phi/2}\hat{\tilde c}^\dagger_{lg}$}. When driving close to resonance of both transitions, it is convenient to go to a rotating-gauge frame (RGF) via
\begin{align}
    \hat{\tilde c}^\dagger_{le}\to\im^le^{\im t\left[(\omega_0-\omega_1)l+\omega_0\right]}\hat{a}^\dagger_{le},~\hat{\tilde c}^\dagger_{lg}\to\im^le^{\im t(\omega_0-\omega_1)l}\hat{a}^\dagger_{lg}.
\end{align} \Rev{Under this unitary transformation we can identify fast-rotating terms which can be removed since they average out. To avoid exciting the atoms to undesired state during laser interrogation, we assume $|\tilde\Omega_0|,~|\tilde\Omega_1|\ll\Delta$. This leads us to a tilted RM (t-RM) model: } 
\begin{align}
    \hat H_\text{t-RM}/\hbar = \hat H_\text{RM}/\hbar +\delta_t\sum_{l,\alpha=e/g}l\hat{a}^\dagger_{l\alpha}\hat a_{l\alpha},
\end{align} where
\begin{align}
    \hat H_\text{RM}/\hbar=&\sum_l\left(\frac{\Omega_A}{2}\hat{a}^\dagger_{le}\hat a_{lg}+\frac{\Omega_B}{2}\hat{a}^\dagger_{le}\hat a_{l+1g}+\text{h.c.}\right)\nonumber\\
    &+\frac{\delta}{2}\sum_l\left(\hat{a}^\dagger_{lg}\hat a_{lg}-\hat{a}^\dagger_{le}\hat a_{le}\right),
    \label{eq:rm}
\end{align}
with \Rev{$\Omega_A=\tilde\Omega_0,~\Omega_B=-\im\tilde\Omega_1$, both real-valued}. When both drives are on resonance ($\delta=\delta_t=0$), \Rev{we obtain} the SSH model: $\hat H_\text{SSH}/\hbar=\sum_l\left(\Omega_A\hat{a}^\dagger_{le}\hat a_{lg}+\Omega_B\hat{a}^\dagger_{le}\hat a_{l+1g}+\text{h.c.}\right)/2$\Rev{. At this leading order, we have ignored the AC Stark shifts generated by both lasers, see \apdx{acstark} for details. These terms contribute to undesired noise that we include in later sections.}

A schematic visualization \Rev{of the above protocol} is shown in \fig{sys}-(a). All of the model parameters, namely, $\Omega_A,~\Omega_B,~\delta$ and $\delta_t$ can be tuned: $\delta$ and $\delta_t$ can be tuned via tuning laser frequencies as long as they are smaller than $\Delta$; $\Omega_A$ and $\Omega_B$ can be tuned either by tuning the laser Rabi frequencies $\Omega^c$ and $\Omega^s$, the lattice depth (which modifies $I_0$ and $J$), the spin-orbit coupling phase $\phi$\cite{Simon2023}, or the tilting potential $\hbar\Delta$\cite{zheng22nat}. 

\begin{figure}
    \centering
    \includegraphics[width=\linewidth]{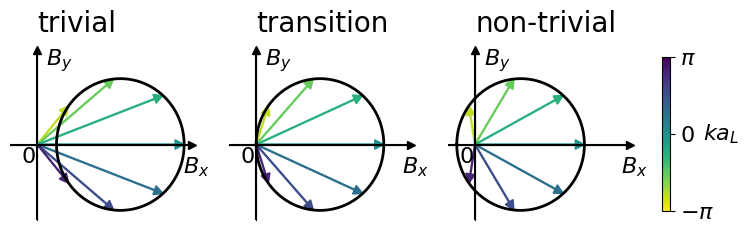}
    \caption{\Rev{Schematic illustration of the winding number of the SSH model. In all panels, the arrows display the effective magnetic field of the SSH model, $\vec B(k)$, as $ka_L$ varies over the BZ. Different colors represent different values of $k$. When $r<1$ (left), the winding number is zero and the system is topologically trivial. At  $r=1$ the system reaches a critical point (center), and for $r>1$ (right) there is a net finite winding number thus a none-trivial topology. Black circles indicate the trajectories traced by the tip of $\vec B(k)$ as $k$ varies across the BZ.}}
    \label{fig:wnum}
\end{figure}

The SSH/RM model has two dimerized phases determined by $r\equiv\Rev{\abs{\Omega_B/\Omega_A}}$. The transition between them is at $r=1$, which sets the topological critical point  for the SSH model. When $r<1$, the SSH model is in the topologically trivial (T) phase, while when $r>1$, the topologically non-trivial (NT) phase. One way to visualize these two phases is by rewriting the model in the quasi-momentum ($k$) basis: $\hat{a}_{k\alpha}=\frac{1}{\sqrt{L}}\sum_le^{\im lka_L}\hat a_{l\alpha}$, which can then be used to define spin operators acting in quasi-momentum space, $\hat S^+_k=\hat{a}^\dagger_{ke}\hat a_{kg}$ and $\hat S^z_k=(\hat{a}^\dagger_{ke}\hat a_{ke}-\hat{a}^\dagger_{kg}\hat a_{kg})/2$. The spin operators satisfy standard commutation relations. In this way, $\hat H_\text{SSH}$ can be written as a spin model
\begin{align}
    \hat H_\text{SSH}=\sum_k\vec{B}(k)\cdot \hat{\vec{S}}_k,
\end{align} 
where $\hat{\vec{S}}_k=(\hat S^x_k,\hat S^y_k,\hat S^z_k)$, and the effective magnetic field is defined as $\vec{B}(k)\equiv(\Omega_A+\Omega_B\cos ka_L,\Omega_B\sin ka_L,0)\equiv(\abs{B}\cos(\phi_k),\abs{B}\sin(\phi_k),0)$ \Rev{, where $\phi_k$ is the angle between $\vec{B}(k)$ and the $x$-axis in the $x-y$ plane of the Bloch sphere}. The topology of the SSH model can thus be seen through a closed trajectory of $\vec{B}$ along the Brillouin zone (BZ), $ka_L\in (-\pi,\pi]$. \Rev{As shown in \fig{wnum}, t}he T/NT phase depends on whether or not this trajectory winds around the origin, as defined by the ``winding number'' $\mathcal{W}$ or the Zak phase $\phi_\text{Zak}=-\pi\mathcal{W}$. Mathematically, for the SSH model, $\mathcal{W}_\text{SSH}=-\frac{1}{\pi}\int_\text{BZ}\mathcal{A}(k)dk$ with the Berry phase $\mathcal{A}(k)=-\frac12d\phi_k/dk$\Rev{, resulting in} $\mathcal{W}_\text{SSH}=0$ when $r<1$ and $\mathcal{W}_\text{SSH}=1$ when $r>1$.

A non-trivial Berry phase can give rise to topological Thouless pumping\cite{thouless83prb,niu1984jp} when one varies the RM model parameters ($\Omega_A,~\Omega_B$ and $\delta$) in a way that they adiabatically return to their initial values after a pump cycle time $\tau$. \Rev{We discussed this in the latter part of this manuscript, \sect{topomwi} and in particular, in \fig{pump} and corresponding discussions}. The particle transport at the end of each cycle is restricted to be an integer number set by $\mathcal{W}$ as the trajectory of $(\delta,\Omega_A-\Omega_B)$ winds around the origin once, otherwise there is no particle transport. The direction towards which the particle moves is determined by its initial state and the winding direction of $(\delta,\Omega_A-\Omega_B)$. \Rev{The topological Thouless pumping has been shown to be robust against weak perturbations including certain interactions, time-independent spatial disorder and time-dependent disorder.\cite{niu1984jp, xiao10rmp, song19prb, citro23natrev, walter23natphys}}

\section{A spectroscopic probe of the topological phase transition}\label{pt}
In the following section, we discuss how we can use standard clock spectroscopy to probe \Rev{$\mathcal{W}_\text{SSH}$,} the topological nature of the SPT phases in our system, either when $\delta=\delta_t=0$, or when $\delta_t=0$ with fixed $\delta$.

We use the states $\{\ket{\downarrow}_l\equiv\ket{l+1,g},\ket{\uparrow}_l\equiv \ket{l,e}\}$ as an effective two-level system to perform the read out, as done in prior work\cite{aeppli22tiltSr,bothwell2022}. In terms of these states, we can define the corresponding sideband spin-$1/2$ operators: $\hat I^y_l\equiv(\hat{a}^\dagger_{le}\hat a_{l+1g}-\hat a_{l+1g}^\dagger\hat{a}_{le})/(2\im)$, $\hat I^x_l\equiv(\hat{a}^\dagger_{le}\hat a_{l+1g}+\hat a_{l+1g}^\dagger\hat{a}_{le})/2$ and $\hat I^z_l\equiv(\hat{a}^\dagger_{le}\hat a_{le}-\hat a_{l+1g}^\dagger \hat{a}_{l+1g} )/2$, which satisfy the appropriate commutation relations. In a typical clock measurement, one measures the \emph{global} observables given by the sum of all local observables, namely $\hat O=\sum_l\hat O_l$. We also simplify the expectation value of an operator as $O=\braket{\hat O}$. Additionally, since $\hat I_z$ is the global particle number difference between $e$ and $g$ state\Rev{s}, $\hat I_z=\hat S_z\equiv\sum_l(\hat{n}_{le}-\hat{n}_{lg})/2$, where $\hat{n}_{l,e/g}=\hat{a}_{l,e/g}^\dagger\hat a_{l,e/g}$.

\begin{figure*}
    \centering
    \includegraphics[width=0.8\linewidth]{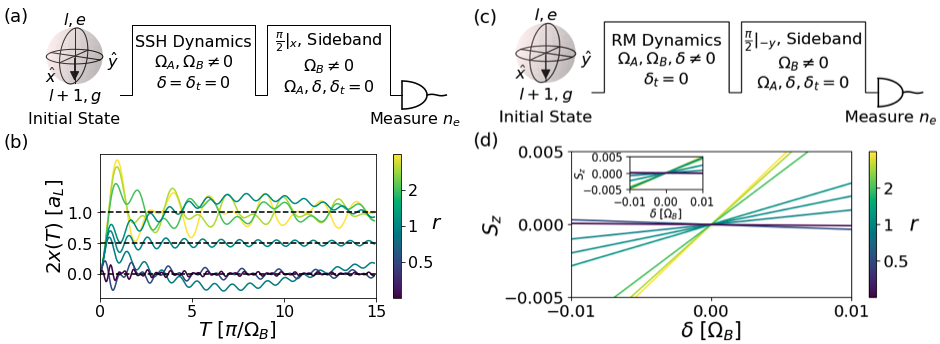}
    \caption{Probing topological phases in the SSH model. (a)(c): Protocols in the rotating-gauge frame. (b)(d): numerical simulations. (a)(b): Winding number measurements with $\delta=\delta_t=0$, compared to their expected values ($0$, $0.5$ and $1$, shown as black horizontal dashed lines) in their corresponding topological phases. \Rev{Line colors correspond} to different $r$ values indicated by the colorbar. (c)(d): One-step winding number measurements with $\delta\neq0$ and $\delta_t=0$. (d): One-step measurement results as a function of $\delta/\Omega_B$ at $\Omega_Bt=\pi$ (main panel) and $\Omega_Bt=15\pi$ (inset). \Rev{At long time, t}he slope at $\delta=0$ gives half of the winding number value. The main panel and the inset share the same $x-$ and $y-$axis, with parameters: $\delta/\Omega_B=-0.1\cdots0.1$, $r=0.3\cdots3$ and $\delta_t=0$.}
    \label{fig:phase}
\end{figure*}

We sketch the schematics of \Rev{the $\mathcal{W}_\text{SSH}$ measurement protocol} in \fig{phase}-(a): first, we prepare the atoms in the $g$ internal level, each in a Wannier-Stark eigenstate. Similar to Rabi spectroscopy, we illuminate the atoms for time $t$. However, in our case, to implement the SSH model, we turn both laser tones on, and drive the carrier and sideband transitions \emph{simultaneously} on resonance, namely, $\delta=\delta_t=0$. At the end of the time evolution, instead of measuring the excitation fraction directly, we need to measure the total sideband coherence, $I_y$, along a quadrature perpendicular to the already-applied sideband tone. The latter can be measured by suddenly turning off the carrier drive while keeping the sideband drive on, for the specific time required to drive a sideband $\pi/2-$pulse, $U^{m_1}_\text{RGF}=\exp{\left[-\im\frac{\pi}{2}\left(\sum_l \hat I^x_l\right)\right]}$, followed by measuring total excited and ground state populations $n_{e/g}\equiv\braket{\hat n_{e/g}}=\sum_l\braket{\hat n_{l,e/g}}$ \Rev{to obtain} $S_z$.

By measuring the sideband coherence $I_y$, we can in fact measure the current \Rev{in the SSH chain} given by the rate of change of the wavefunction displacement:
\begin{align}
    \frac{\dot{\hat{x}}}{a_L}=\Omega_B\hat I_y,
\end{align} where the displacement is $\hat x= a_L\sum_ll(\hat n_{le}+\hat n_{lg})$. A similar protocol was performed in quantum gas microscopes to measure currents with single-site resolution\cite{imperto24prl}.

To probe the topological phase transition, one has to repeat this protocol and measure $I_y(t)$ at different time $t$ to obtain its time integral. When weighted by $\Omega_B$, we obtain the displacement of the atomic wavefunction from its initial location, $x(T)$:
\begin{align}
   \frac{x(T)}{a_L}=\Omega_B\int_0^TdtI_y(t).
\end{align} 
We show in \fig{phase}-(b) that $x(T)$ can be used to probe topological phases of the SSH model, i.e.,
\begin{align}
  \frac{x(T)}{a_L}=\frac{\mathcal{W}_\text{SSH}}{2}+\text{oscillating terms}.
\end{align} In fact, the quantity $x(T)$ is equivalent to the ``mean displacement''(MD), a bulk observable that can be used as a marker of topological phases\cite{rudner09prl,cardano17natcomm,maffei18njp}. The MD has been used to measure winding numbers in twisted photons\cite{cardano17natcomm}, and can be used to study the topological properties of more complicated models\cite{maffei18njp}. 

The simple limiting cases when the system is totally dimerized, with $\Omega_B\neq0,~\Omega_A=0$ and  $\Omega_A\neq0,~\Omega_B=0$, can serve to illustrate the distinct behaviors in the two different topological phases. In the former case\Rev{, the NT phase,} the atom performs standard Rabi oscillations in the sideband transition $I_y(t)= \sin(\Omega_B t)/2$ and therefore $x(T)/a_L=(1-\cos(\Omega_B T))/2$; while in the latter case, \Rev{the T phase}, only the carrier transition is driven, thus atoms remain localized at their initial sites and thus $I_y(t)=x(T)=0$.

\Rev{In the presence of a small detuning, $\delta$, from the carrier transition}, we can measure the MD in \emph{one step}, without integrating over time, by measuring the $x$-component of the sideband coherence $I_x$ after time evolution. We measure $I_x$ by suddenly applying a phase jump of $-\pi/2$ \Rev{to} the Rabi sideband drive while simultaneously turning off the carrier Rabi drive, for a time necessary to realize a $\pi/2$-pulse,  $U^{m_2}_\text{RGF}=\exp{\left[\im\frac{\pi}{2}\left(\sum_l \hat I^y_l\right)\right]}$, followed by a measurement of the excited/ground state populations $n_{e,g}$. We display the above protocol in \fig{phase}-(c). One can show analytically using linear responses that when $\delta_t,\delta\ll\Omega_A,\Omega_B$, 
\begin{align}
    I_x(t)\simeq\frac{\delta}{\Omega_B}\frac{\mathcal{W_\text{SSH}}}{2}+\delta_t\tilde S(\Omega_A,\Omega_B)+\text{osc. terms},
    \label{eq:Ix}
\end{align}where $\tilde S(\Omega_A,\Omega_B)$ is a real-valued function not directly related to $\mathcal{W_\text{SSH}}$ (see \apdx{Ixlr} and \fig{sfcn} for details). The above equation \Rev{indicates} that the kinetic term $I_x$\cite{imperto24prl} responds linearly to the inversion-symmetry-breaking terms $\delta$ and $\delta_t$, with a slope related to $\mathcal{W}_\text{SSH}$ for the carrier detuning $\delta$. Again, in the simple limiting case when the system is fully dimerized, we see that $I_x(t)=0$ when $\Omega_A\neq0,~\Omega_B=0$, while $I_x(t)\simeq(\delta+\delta_t)(1-\cos(\Omega_Bt))/2$ when $\Omega_A=0,~\Omega_B\neq0$. 

We show the numerical simulations of $I_x(t)$ with $\delta_t=0$, at $t=\pi/\Omega_B$ and $t=15\pi/\Omega_B$ in \fig{phase}-(d) \Rev{and its inset}. This provides us a one-step measurement of $\mathcal{W}_\text{SSH}$: as long as we know the values of $\Omega_B$ and $\delta$, we can measure $\mathcal{W}_\text{SSH}$ via $I_x$. \Rev{In fact, $I_x$ is} a band correlation function discussed in prior work\cite{song19prb}\Rev{, and has been found  to be robust to certain types of disorder.} 

\Rev{The same quantity can alternatively be measured by first preparing an initial state $(\ket{0,g}+\ket{-1,e})/\sqrt{2}$, then turning on the RM dynamics for time $t$, followed by measuring $-S_z$. We show this protocol in \apdx{sshspec2}. Mathematically, this protocol is equivalent to the protocol discussed above.}

\Rev{It is worth noting that \eq{Ix} assumes an infinite number of lattice sites with periodic boundary conditions. In \apdx{tptLs}, we discuss how finite-size effects and open boundary conditions modify $I_x(t=15\pi/\Omega_B)$ and affect the sharpness of the phase transition. We also discuss the robustness of $I_x(t=15\pi/\Omega_B)$ to various levels of static, shot-to-shot amplitude uncertainties of $\Omega_A$ and $\Omega_B$ in finite  chains.}

\section{SSH clock spectroscopy}\label{sshspec}

We now discuss a clock spectroscopy protocol that is robust against unwanted noise in laser parameters thanks to the underlying topology of the system. \Rev{We focus on time-independent global and local amplitude laser noise. The global amplitude noise is modeled as global variations of the Rabi frequency. On the other hand, the local amplitude noise is modeled as a position-dependent Rabi frequency variations felt by each atom as it spreads over the radial direction of the OLC due to their finite radial temperature\cite{blatt09pra,kim23prl}. We do not include global laser phase noise in our analysis, because clock spectroscopy is inherently sensitive to it and thus such noise is always detrimental. Additionally, The SSH model is not immune to such noise as it breaks the symmetry of the model (i.e. the ``Chiral symmetry'').}

In conventional Rabi spectroscopy, the atomic frequency is inferred via the so called Rabi lineshape \Rev{(illustrated in the inset of the upper panel of \fig{sys}-(b))}, obtained by first driving a carrier laser \Rev{with Rabi frequency $\Omega_R$ for a time $t_R$ such that $\Omega_R t_\pi^R=\pi$}, with \Rev{the laser frequency} detuned from the clock transition frequency by a detuning $\delta_L=\omega_0-\omega_a$, followed by measuring populations $n_e,n_g$ as a function of $\delta_L$. The resonance value $\omega_a=\omega_0$ is given by the solution of $(n^{-1}_e(\delta_c)+n^{-1}_e(\delta'_c))/2=\delta_{resonance}=0$, typically by choosing $\delta_c$ and $\delta'_c$ at the positive and negative slopes of the Rabi line shape at half maximum (FWHM), namely, where the line shape achieves maximum positive and negative slopes. \Rev{Uncertainties in $\Omega_R$ lead to an uncertainty of $t^R_\pi$ that in turn induces both a systematic error on the inferred resonant detuning at the FWHM, as well as a reduction of the signal at $\pm\delta_c$.}

Our SSH clock spectroscopy adopts a similar idea as the Rabi spectroscopy, but we determine $\omega_a$ using the winding number measurement protocol, $I_x$, as discussed in \sect{pt} and \fig{phase}-(c),(d). Specifically, we operate in the NT phase by setting $\Omega_B>\Omega_A$, let the system evolve for a fixed time \Rev{$t^B_\pi =5\pi/\Omega_B$ when the value of $I_x$ saturates around $\mathcal{W}/2$}, then measure $I_x$.% ( See \apdx{specsens}). 

\begin{figure}
    \centering
    \includegraphics[width=0.9\linewidth]{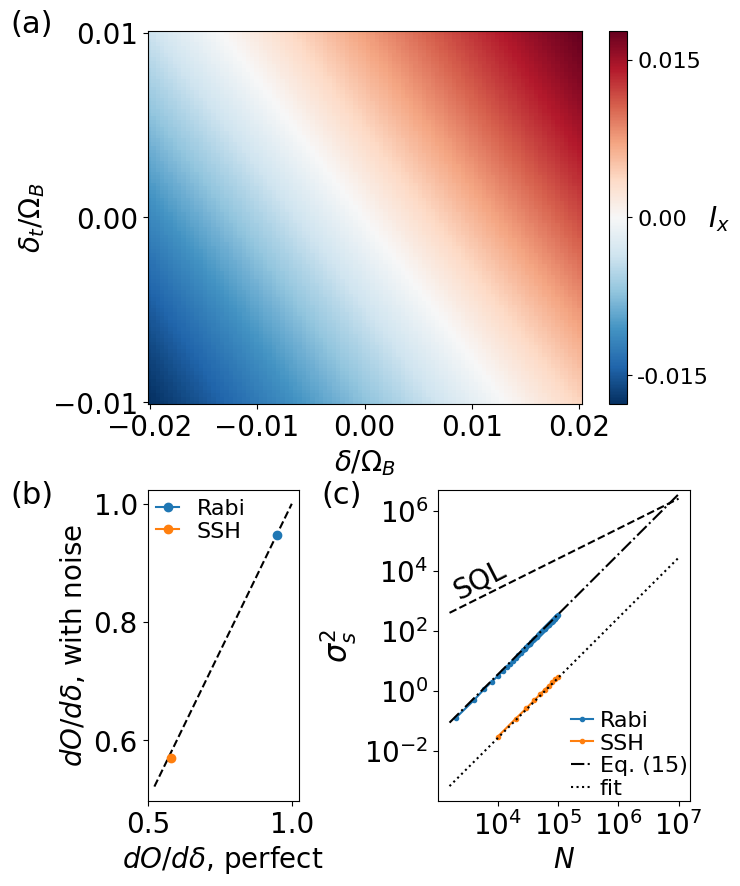}
    \caption{(a): The signal of the SSH spectroscopy, $I_x$, at different $\delta$ and $\delta_t$ values without any experimental imperfections. (b) and (c): Comparisons between Rabi spectroscopy (in blue) and SSH spectroscopy (in orange) with $\Omega_A/(2.2 \pi)=\Omega_B/(2 \pi)=  10$ Hz, over \Rev{$200$} realizations of \Rev{global noise and $N$ atom-dependent noise as detailed in the the main text.} (b): \Rev{How signal varies with respect to $\delta$, $dO/d\delta$, at operation points $\delta_c$'s of the two protocols. For Rabi spectroscopy, $O=S_z$ and $\delta_c\sim0.8\Omega$; for SSH spectroscopy, $O=I_x$ and $\delta_c=0$, averaged over all noise realizations.} (c): $N-$dependence of $\sigma^2_s$ \Rev{of the two protocols. The dashed line shows the fundamental sensitivity achievable for uncorrelated atoms, the SQL. The dashed-dotted line compares Rabi results to \eq{rabisens}. The dotted line is a fit to the SSH results, which is $y=\gamma^2\sigma_a^2N^2$, where $\gamma^2=0.00027$.}}
    \label{fig:clock}
\end{figure}

In addition to the carrier detuning, however, in our symmetry-protected SSH clock, there is also \Rev{an additional detuning $\delta_t$}. In \fig{clock}-(a) we show that \Rev{$I_x$} varies linearly with respect to $\delta$ and $\delta_t$, as discussed in \eq{Ix}. The relevant bare parameters, $\omega_a$ and $\Delta$, are inferred by finding a set of detunings that satisfy $I_x(\delta,\delta_t)+I_x(-\delta,-\delta_t)=0$. \Rev{To measure clock transition frequency, we set $\delta_t=0$, in which case the frequency difference between the two laser tones is $\Delta$.}

\Rev{In the presence the noise discussed above,  the SSH Hamiltonian becomes:
$\hat H_\text{SSH}/\hbar = \sum_{il}\left(\Omega^i_A\hat{a}^\dagger_{ile}\hat a_{ilg}+\Omega^i_B\hat{a}^\dagger_{ile}\hat a_{il+1g}+\text{h.c.}\right)/2$, where the Rabi frequencies applied to each atom $i$ is
\begin{align}
    \Omega^i_{A/B}=\bar{\Omega}_{A/B}\left(1+\epsilon_a\right)\left(1+\epsilon_i\right),
\end{align} 
with $\bar{\Omega}_{A/B}$ the ideal value, $\epsilon_a$ the shot-to-shot global amplitude noise, and $\epsilon_i$ the local amplitude noise. We take $\epsilon_a$ and $\epsilon_i$ to be zero-mean Gaussian random variables with standard deviation $\sigma_a$ and $\sigma_i$, respectively. We also include the additional AC Stark shifts (see\apdx{acstark}). In our simulation, we use current experimentally-relevant values and set $\sigma_a=0.001$ and $\sigma_a=0.04$\cite{yan25,blatt09pra,kim23prl}.} We also set the Rabi frequenc\Rev{ies} to be $\Omega_B/(2\pi)=10$Hz \Rev{and $\Omega_A = \Omega_B/1.1$}.

We compare the signals of \Rev{the two protocols, i.e.
$dO/d\delta$ with $O=S_z$ for Rabi and $O=I_x$ for SSH,} at their respective operation points $\delta_c$, with $\Omega_R=\Omega_B$. \Rev{We observe minimal signal reductions in both protocols as shown in \fig{clock}-(b).}

To quantify the robustness of a measurement protocol against imperfections, we calculate the clock sensitivity to $\delta$, \Rev{for} a system of $N$ non-interacting atoms \Rev{as}:
\begin{align}
    \Delta^2\delta =  \frac{\braket {\Delta ^2 \hat O}}{t^2_L(d\braket{\hat O}/d\varphi)^2}\Rev{\bigg|_{\delta\to\delta_c}},~~\braket{\Delta^2 \hat O}=N/4+\sigma^2_s,
    \label{eq:sensitivity}
\end{align}where $t_L$ is the laser interrogation time, $\varphi$ is the accumulated phase due to $\delta$\Rev{, $\delta_c$ is the working point of the noise-free protocol,} $\hat O$ is the measured observable which gives the signal $\braket{\hat O}$, \Rev{and associated} variance $\braket{\Delta ^2 \hat O}$. The variance includes both the quantum noise, or SQL, of $N/4$ for non-interacting atoms, and the statistical noise $\sigma^2_s$, coming from the sensitivity of a protocol to technical noise. For the collective observable measured in the described protocols, $\hat O=\hat S_z$, the global noise increases quadratically in $N$\cite{lucke14prl,gabardos20prl}, namely,
\begin{align}
    \sigma^2_{s,gl}\simeq \sum_{\beta} \sigma^2_{\Omega_\beta,gl}\underbrace{\sum_{i\neq j}\left(\frac{\partial \braket{\hat O_i}}{\partial{\Omega_\beta}}\Big |_{\bar{\Omega}_\beta}\right)\left(\frac{\partial \braket{\hat O_j}}{\partial{\Omega_\beta}}\Big |_{\bar{\Omega}_\beta}\right)}_{\propto N^2},
\end{align} where $gl$ stands for global, $\beta$ is all the Rabi frequencies involved in the protocol, i.e., $\beta=R$ for the Rabi spectroscopy and $\beta=A,B$ for the SSH spectroscopy. \Rev{For Rabi spectroscopy, it can be analytically calculated that 
\begin{align}
    \sigma^2_{s,gl}\simeq0.035N^2\sigma^2_{\Omega_R,gl}.
    \label{eq:rabisens}
\end{align}
The noise affecting individual atoms, on the other hand, scales as $\sigma^2_{s,ind}\propto N$, and is suppressed when $N$ is large.}

In the case of the SSH spectroscopy, $\sigma^2_s$ is suppressed \Rev{since $\partial\braket{\hat I^x_i}/\partial{\bf\Omega}$ is small, as} $I_x\propto\mathcal{W}_\text{SSH}\delta$ where $\mathcal{W}_\text{SSH}$ is topologically protected against static amplitude uncertainties of $\Omega_{A,B}$\Rev{\cite{meier18sci,song19prb}. We discuss this briefly in \apdx{tptLs}.}

We compare $\sigma^2_s$ of our protocol with that of the Rabi spectroscopy in \fig{clock}-(c) as a function of \Rev{total particle number $N$}, assuming perfect $I_x$ measurements. We observe \Rev{a reduction of $\sigma^2_s$ in the SSH spectroscopy. By fitting $\sigma^2_s$ at different $N$ values to $y=\gamma^2\sigma^2N^2$, we obtain $\gamma^2\simeq0.00027$.} \Rev{We also compare our simulation with the SQL shown by the black dashed line in \fig{clock}-(c) and compare $\sigma^2_s$ of the Rabi spectroscopy to \eq{rabisens}. Given the small Rabi frequency noise we consider, $\sigma^2_s$ of the Rabi spectroscopy becomes comparable with the SQL only for $N\gtrsim10^7$. That means that while not so relevant under current conditions, the discussed SSH protocol could be beneficial to future generation OLCs capable to interrogate larger atom arrays. Other types of local amplitude noise ignored here could push the utility of our protocol even to lower $N$.}

If it is not possible to measure $I_x$ perfectly, the same quantity can alternatively be measured \Rev{using the protocol mentioned in the end of the last section, or in \apdx{sshspec2}, where the initial state can be prepared adiabatically, and $S_z$ can be measured fairly accurately.}

%Last but not least, since the laser interrogation time of our protocol is $5$ times that of the Rabi protocol, a fair comparison of the two protocols is to compare one SSH measurement to $5$ Rabi measurements. We compare these two protocols in \apdx{specsens}.

%\Rev{Last but not least, we would like to reiterate that the SSH spectroscopy is expected to be sensitive to perturbations that break its symmetry, such as the global laser \emph{phase} noise. Moreover, a sensitivity to the laser phase is anyway required for any clock spectroscopy protocol.}

\begin{figure*}
  \centering
    \includegraphics[width=\linewidth]{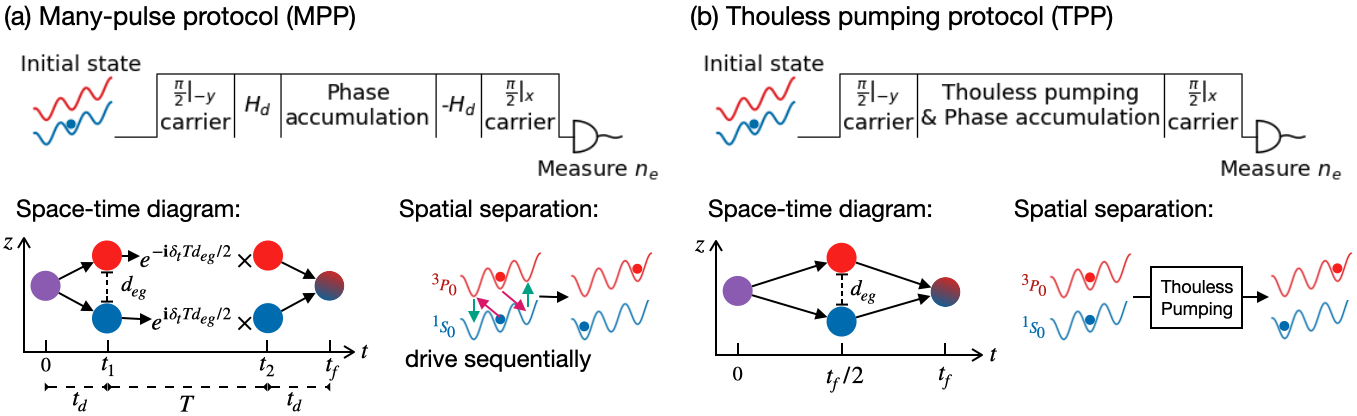}
    \setlength{\abovecaptionskip}{-5pt}
    \setlength{\belowcaptionskip}{-5pt}
    \caption{\Rev{MWIs in OLCs. (a) The ``many-pulse protocol'' (MPP). (b) The ``Thouless pumping protocol'' (TPP).  Upper panels: the protocol sequence of each protocol as described in the main text. Lower left panels: the space-time diagram of each protocol. Lower right panels: pictorial indication of one drive cycle of each protocol that creates $d_{eg}=2a_L$. }}
   \label{fig:interf}
\end{figure*}

\section{Topologically-pumped matter-wave interferometer}\label{topomwi}
%\tx{Need to edit with new pumps. Interesting: TPPs have built-in echos actually. Not going to include this but I do want to comment on old pumps: overall smaller drive, errors can be reduced, to some extent, with echos. Basically: circle pumps easier to scale $\tau$, good for constant-in-time scans, but old pumps can be good due to smaller driving amplitudes.}

Conventional interferometers detect force fields, such as the ones proportional to the local gravitational acceleration $g$\cite{peters99nat} and the Newtonian gravitational constant $G$\cite{rosi14nat}, via measuring the differential phase, $\varphi$, experienced by matter waves traveling through different paths in the force field. \Rev{These MWIs typically }operate using free-falling atoms that enjoy a phase accumulation that scales quadratic in time \cite{peters99nat}. \Rev{While} trapping the atoms in optical lattices \cite{clade05epl,carriere12pra,guglielmo16pra,xu19sci,panda24natphys,panda24nat} has the advantage of a much longer interrogation time\Rev{, currently} approaching minutes, the \Rev{achievable} phase accumulation rate is limited to scaling linearly with time. A way to further enhance the accumulated phase is via the application of multiple pulses\cite{Muller2008,pelle13pra} that helps to increase the \Rev{spatial} separation between the two parts of the wavepacket.  

In a possible implementation of the protocol\cite{pelle13pra,chu21prl}, one first applies an initial carrier $\pi/2$-pulse on atoms prepared in their ground state, generating a coherent superposition \Rev{on} a single site $l$: $\ket{\psi^{MWI}_0}=(\ket{l,g}+\ket{l,e})/\sqrt{2}$. Then, \Rev{for a duration $t_d$, }a sequence of $N^M_p$ composite $\pi$-pulses, consisting of a sideband drive followed by a carrier drive, denoted as $H_d$, is applied to induce a spatial separation between $\ket{e}$ and $\ket{g}$ states: $d_{eg}=2N^M_pa_L$. Afterwards, within a dark time $T\gg t_d$, the \Rev{two states} accumulate a differential phase since \Rev{they are spatially} separated \Rev{at} locations with different gravitational potential. Then, \Rev{after a pulse sequence that reverses the spatial separation, denoted as $-H_d$, the atoms are recombined. Lastly, we measure the accumulated phase difference} by applying another local carrier $\pi/2$-pulse that converts the $y$-component of the carrier coherence into population, which is the quantity measured at the end of the sequence. Assuming fast and perfect $\pi/2$-pulses \Rev{for the state preparation and measurement}, the total time of this protocol is $t_f=2t_d+T$. This MWI signal depends linearly on the accumulated phase. We refer to this protocol as the ``many-pulse'' protocol (MPP), and illustrate it in \fig{interf}-(a). Assuming perfect drives, the ideal accumulated phase of the MPP is 
\begin{align}
    \varphi^M_\text{ideal}=\delta_tTd_{eg}/a_L=2\delta_t(t_f-2t_d)N^M_p.
    \label{eq:phimpp}
\end{align}

\begin{figure}
  \centering
    \includegraphics[width=\linewidth]{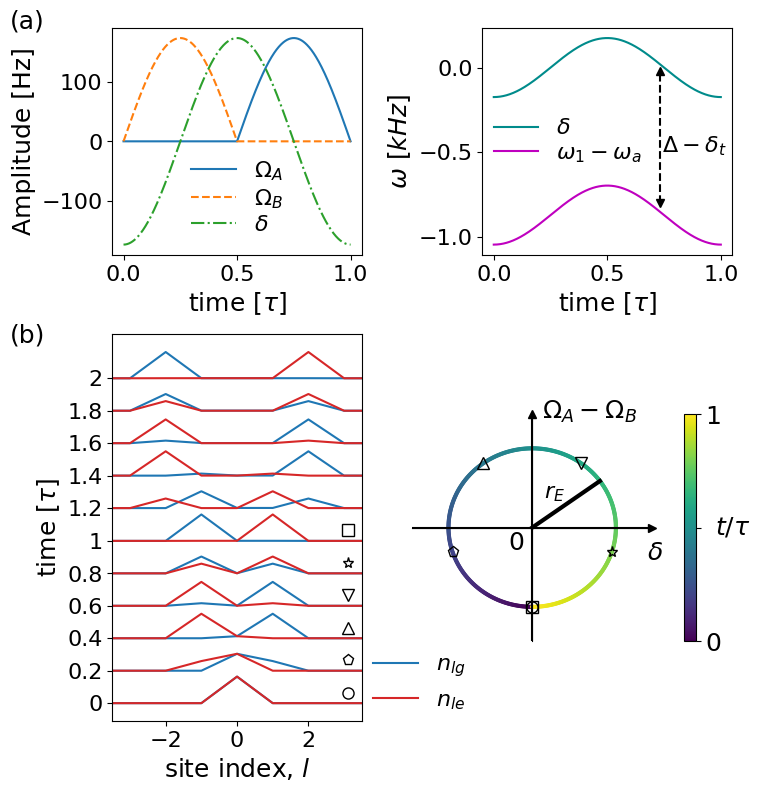}
    \setlength{\abovecaptionskip}{-5pt}
    \setlength{\belowcaptionskip}{-5pt}
    \caption{\Rev{Thouless pumping in a tilted 1D OLC. (a): Time-dependence of the Hamiltonian parameters $\Omega_A,~\Omega_B,$ and $\delta$ (left panel) and laser frequencies (right panel) of one Thouless pumping cycle of time $\tau=1/12~$second. (b): Left panel: the dynamics of the probability density function of the $e$ (red) and $g$ (blue) states of an atom within two Thouless pumping cycle, with initial state: $\left(\ket{0,g}+\ket{0,e}\right)/\sqrt{2}$. Right panel: time-dependence of $(\delta,~\Omega_A-\Omega_B)$ of one Thouless pumping cycle, with time indicated by the colormap. Open symbols on the left panel corresponds to the time and Hamiltonian parameters shown on the right panel. Note that the first symbol (open circle) and the last symbol (open square) are on top of each other, indicating the time-periodicity of Thouless pumping cycles.}}
   \label{fig:pump}
\end{figure}

We now discuss an alternative MWI protocol that shows reduced sensitivity of unwanted noise thanks to the use of topologically-protected adiabatic transfer instead of a multi-pulse sequence. The basic idea is to achieve the desired spatial separation using the topological ``Thouless pumping'' protocol (TPP), shown in \fig{interf}-(b). In this case, within the same time duration $t_f$, half of the time is spent to repeat $N^T_p$ Thouless pumping cycles with a cycle time $\tau$, in order to spatially separate the $\ket{g}$ and $\ket{e}$ states by $d_{eg}=2N^T_pa_L$. Then, the same amount of time is used to bring the two states back to the initial location, and measure the accumulated phase in the same way as in the MPP. No dark time is needed in this protocol as the two states continuously accumulate differential phase during the adiabatic Thouless pumping cycles. We sketch in \Rev{\fig{pump}-(a)} the time variation of the RM parameters \Rev{(left panel)} used to perform a pumping cycle with duration $\tau=1/12$ seconds and their corresponding laser frequencies \Rev{(right panel)}: in the protocol, we \Rev{vary} $\omega_1$ together with $\omega_0$, so that the value of $\delta_t$ is kept at a fixed value at all time.  \Rev{In the left panel of \fig{pump}-(b), we show the spatial distribution of the states $e$ and $g$ in the lattice, $n_{l,e/g}=\braket{\hat a^\dagger_{l,e/g}\hat a_{l,e/g}}$ during two Thouless pumping cycles. We also show in the right panel of \fig{pump}-(b) the time-dependence of $(\delta,~\Omega_A-\Omega_B)$ within one Thouless pumping cycle.}

Contrary to the MPP, the ideal accumulated phase scales \emph{quadratically} with $t_f$\Rev{, which is is also the case for free-falling atoms}:
\begin{align}
    \varphi^T_\text{ideal}=4\delta_t\sum_{n=0}^{N^T_p} (n\tau)=2\delta_t\tau\left(N^T_p\right)^2 = \delta_tt^2_f/(2\tau).
    \label{eq:phitpp}
\end{align}

We compare the MPPs and TPP performance with fixed $t_f=4s$ and $\varphi^M_\text{ideal}=\varphi^T_\text{ideal}$. In particular, we consider three different MPPs, i.e., protocols $P_0$ to $P_2$ with $P_0$: $\Omega/( 2\pi) = 24$ Hz,  $N^M_p=24$, $T=2$ s; \Rev{$P_1$: $\Omega/(2\pi)= 37.5$ Hz, $N^M_p=15$, $T=3.2$ s; and $P_2$: $\Omega/(2\pi)=39.86$ Hz, $N^M_p=65$, $T\simeq0.74$ s,} adjusting the laser power to achieve the same Rabi frequency $\Omega$ for both the carrier and the sideband drives. \Rev{These three chosen protocols aim to capture different ways to improve the MWI sequence by maximizing phase accumulations, although by no means they are  the optimal ones: $P_1$ has long dark time $T$, and $P_2$ has long $t_d$ thus leads to large $d_{eg}$, while $P_0$ balances $t_d$ and $T$.} We consider one TPP with $\tau=1/12$ s and $N^T_p=24$ (protocol $P^{\tau=1/12~\rm{s}}_{\rm TPP}$). All of the protocols \Rev{are} designed to achieve the same phase accumulation under ideal conditions.

To investigate the robustness of these protocols, we account for relevant sources of noise, described by the function \Rev{$\Omega^i_{A/B}=\bar{\Omega}_{A/B}\left(1+\epsilon_a\right)\left(1+\epsilon_i\right)$, the same as the previous section. We also include AC Stark shifts as discussed in \apdx{acstark}. The time dependence of $\delta$ in TPP thus becomes $\delta(t)=\delta_p(t)+\bar\delta_\text{ACS}(t)+\delta^\epsilon_\text{ACS}(t)$, where $\delta_p(t)$ is the actual detuning of the laser frequency $\omega_0$ from the atomic transition frequency $\omega_a$, which we vary over time during the protocol. The quantity $\bar\delta_\text{ACS}(t)$ is the ideal AC Stark shifts that can be calculated based on the instantaneous Rabi frequencies and $\Delta$. In the presence of noise in Rabi frequency amplitudes, $\delta(t)$ has a time-dependent noise given by: $\delta^\epsilon_\text{ACS}=\delta_\text{ACS}(t)-\bar\delta_\text{ACS}(t)\neq0$.}

%Both \eq{phimpp} and \eq{phitpp} indicate that the accumulated phases of the two protocols depend on (1) the achievable spatial separation $d_{eg}$ and (2) the phase accumulation time. Here, we examine how systematic imperfections affect the achievable spatial separation. It is worth noting that, since we consider an atom in a coherent superposition, due to experimental imperfections, it is no longer the case that $n_e=n_g=1/2$ at all times. \Rev{We thus} define the distance between the two states to be:
%\begin{align}
%    d_{eg} = \frac{x_e}{n_e} - \frac{x_g}{n_g}.
%\end{align} 
%Therefore, $d_{eg}$ is a quantity that implies how fast the differential phase accumulates, \emph{solely} due to the spatial separation of $e$ and $g$.

\begin{figure}
  \centering
    \includegraphics[width=\linewidth]{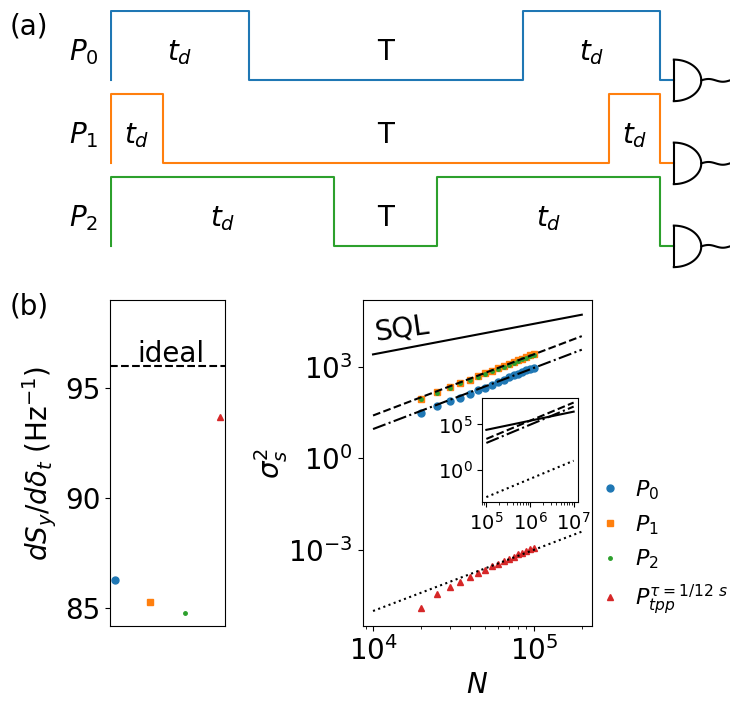}
    \setlength{\abovecaptionskip}{-5pt}
    \setlength{\belowcaptionskip}{-5pt}
    \caption{Comparisons between the MPPs and TPPs \Rev{with} $200$ realizations \Rev{of global imperfections and $N$ non-interacting atoms with atom-dependent imperfections, as discussed in the the main text. (a): Pulse sequences of the MPP protocols $P_{0\cdots2}$. (b): The left panel: how the  signal $S_y$ varies with respect to the detuning $\delta_t$ for the various protocols (symbols) compared to the ideal value (the black dashed line). The right panel: $N$-dependence of $\sigma_s^2$ (symbols), compared with the SQL (black solid line). Black dashed, dashed-dotted and dotted lines: fitting the symbols to $y=\gamma^2\sigma^2_a N^2$, where $\gamma^2\simeq0.25$ (dashed),  $\gamma^2\simeq0.09$ (dashed-dotted) and $\gamma^2\simeq10^{-7}$ (dotted). The inset of the right panel shows the lines of the main panel, extrapolated to larger $N$ values. It shows that for large $N>3 \times 10^6$, the MPPs reach noise levels comparable with the SQL.}}
   \label{fig:interf2}
\end{figure}

\Rev{We investigate the signals obtained for all the protocols and how they change with $\delta_t$, i.e. $dS_y/d\delta_t$. Ideally, these protocols should have the same $dS_y/d\delta_t$ value. However, in the left panel of \fig{interf2}-(b), we observe a reduction in all protocols due to experimental imperfections. Due to the imposed time constraint we also see  a possible lack of full adiabaticity in the TPP protocol. Nevertheless, the TPP still appears to have the largest signal. We would like to note that the lack of full adiabaticity can be resolved by increasing $\tau$, the maximum values of $|\Omega_{A/B}|$}, by optimizing the pulse sequence or via shortcuts-to-adiabaticity protocols\cite{liu2024arxiv}. \Rev{It is worth noting} that for \Rev{topological pumping schemes}, the particle transport is protected even in the presence of small interactions, as long as the interaction strength is small compared to the energy gap of the SSH model.\cite{walter23natphys,xiao10rmp}

\Rev{Lastly, we compare the statistical noise of these protocols. As shown in the right panel of \fig{interf2}-(b), the TPP yields \Rev{significantly} less statistical noise than the MPPs. We also compare all protocols with the SQL and found that the MPPs is comparable with the SQL when $N> 3\times10^6$, while the TPP still remains much smaller than the SQL.} 

%\Rev{In summary, based on} \eq{sensitivity}, for protocols of the same duration and with the same ideal phase accumulation, the TPP has a better sensitivity than the MPPs through both signal enhancement and noise suppression.}

%%%%%%%%%%%%%%%%%%%%%%%%%%%%%%%%%%%%%%%%%%%%
%%%%%%%%%%%%%%%%%%%%%%%%%%%%%%%%%%%%%%%%%%%%
%%%%%%%%%%%%%%%%%%%%%%%%%%%%%%%%%%%%%%%%%%%%

\section{Conclusion and Outlook}
In this manuscript, we described a readily implementable experimental setting to realize a symmetry-protected topological model in a tilted 1D OLC. We discussed how one can measure the system's topological properties, namely the winding number, by taking advantage of the pristine quantum coherence and the exquisite spectral resolution offered by OLCs. We finally discussed two sensing protocols which showed improved sensitivity compared to conventional clock sequences, thanks to the topological robustness of symmetry-protected states against unwanted \Rev{global and local noise present in real experiments. We discussed how topology helps reducing statistical uncertainties, while keeping or even increasing the acquired signal.}

Our work can open up a new path for a new generation of OLCs with topologically-enhanced sensitivity. While we focused here on measuring \Rev{the clock transition frequency $\omega_a$ and the} local gravitational acceleration $g$, our protocols can, in principle, be adapted to improve the measurement resolution of gravitational red shifts\cite{bothwell2022}. Moreover, even though so far we have limited our analysis to non-interacting atoms, an exciting extension is to study how interactions affect the observed topological robustness\cite{lin20prb,walter23natphys}. \Rev{Additionally, even though  we focused our work on static imperfections, our protocols could, in principle, be robust to certain types of time-dependent noise\cite{oelker19nat,li22prapp}.} Finally, generalizing the investigation to more complex systems including higher-dimensional models or systems with more than two internal levels, by incorporating the nuclear spins, will open up an plethora of rich physics, where for the first time the cooperation or competition between interactions and topology can give rise to a new generation of quantum-enhanced and topologically protected sensors.

\section*{Acknowledgments}
We thank Joanna Lis and Maya Miklos for feedback on the manuscript. We also thank Thomas Bilitewski, Mikhail Mamaev, Klaus M\o{}lmer, Bhuvanesh Sundar,  Yanqi Wang and Haoqing Zhang for useful discussions.  We additionally thank Jim McKown, Corey Keasling and Daniel Packman from JILA computing group for supports with relevant hardware used for numerical simulations. This material is based upon work supported by the
SLOAN, the Simons and the Heising-Simons foundations, the Vannevar Bush Faculty Fellowship , the NSF JILA-PFC PHY-2317149 and OMA-2016244 (QLCI), the U.S. Department of Energy, Office of Science, National Quantum Information Science Research Centers, Quantum Systems Accelerator and NIST. TE also acknowledges funding from Swiss National Science Foundation under Advanced grant TMAG-2 209376.

\appendix
\counterwithin{figure}{section}
\section{The SSH Model in a tilted OLC}\label{fullhami}
The Hamiltonian that describes the dynamics of atoms driven by a coherent laser, \Rev{with Rabi frequency $\Omega$, laser frequency $\omega_L$, and wavenumber $k_L$}, is given by
\begin{align}
    H^\Omega_\text{lab}/\hbar=\frac{\Omega}{2}e^{-\im\omega_Lt}\int dxe^{\im k_Lx}\hat\Psi^\dagger_e(x)\hat\Psi_g(x).
\end{align}
Assuming the atoms are trapped in the lowest band of a lattice ,   we can expand the field operator in term of lowest band Wannier states localized at the different lattice sites, the above equation can be rewritten as $\hat\Psi_\alpha=\sum_nw_0(x-na_l)\hat c_{n\alpha}$, where $a_l$ is the lattice constant. We thus get
\begin{align}
    \hat H_\text{lab}=\frac{\Omega}{2}e^{-\im\omega_Lt}\sum_{mn}I_{nm}\hat c^\dagger_{ne}\hat c_{mg},
\end{align}
where
\begin{align}
    I_{nm}=\int dxe^{\im k_Lx}w_0(x-na_l)w_0(x-ma_l)=e^{\im n\phi}I_{m-n},
\end{align}
with $\phi=ka_L$. One can prove that $I_{m-n}=I^*_{n-m}$. If two laser drives are used to illuminate atoms trapped in a tilted system, and $\delta=m-n$, the full Hamiltonian describing the motion and internal dynamics of the atoms is given by
\begin{align}
    &\hat H_\text{lab}/\hbar\nonumber\\
    &=\left(\frac{\Omega^c}{2}e^{-\im\omega_0t}\sum_{l\delta}I_{\delta}e^{\im  l\phi}\hat c^\dagger_{le}\hat c_{l+\delta g}+\text{h.c.}\right)\nonumber\\
    &+\left(\frac{\Omega^s}{2}e^{-\im\omega_1t}\sum_{l\delta}I_{\delta}e^{\im  l\phi}\hat c^\dagger_{le}\hat c_{l+\delta g}+\text{h.c.}\right)\nonumber\\
    &-J\sum_{l\alpha}\hat c^\dagger_{l+1\alpha}c_{l\alpha}+\omega_a\sum_l\hat c^\dagger_{le}\hat c_{le}+\Delta\sum_{l\alpha}l\hat c^\dagger_{l\alpha}\hat c_{l\alpha},
\end{align}
where $\hbar\Delta=Mg_{acc}a_L$ is the the potential energy difference between atoms at adjacent lattice sites. Rewritting the Hamiltonian in terms of Wannier-Stark (WS) states, that diagonalize the motional degrees of freedom: $$\hat H/\hbar=-J\sum_{l\alpha}\hat c^\dagger_{l+1\alpha}\hat c_{l\alpha}+\Delta\sum_{l\alpha}l\hat c^\dagger_{l\alpha}\hat c_{l\alpha}+\text{h.c.},$$ with $\hat c_{l\alpha}=\sum_m\mathcal{J}_{l-m}(2J/\Delta)\hat {\tilde c}_{m\alpha}$, we have\footnote{$\sum_k\mathcal{J}_{\nu+k}(u)\mathcal{J}_{k}(u)e^{\im k\alpha}=\mathcal{J}_{\nu}\left(2u\sin\left(\frac{\alpha}{2}\right)\right)e^{-\im\nu(\pi+\alpha)/2}$}
\begin{align}
    \sum_le^{\im l\phi}\hat c^\dagger_{le}\hat c_{l+\delta g}=\sum_{lk}e^{\im l\phi}e^{-\im(\delta-k)(\pi+\phi)/2}\mathcal J_{\delta-k}(\tilde J)\hat {\tilde c}^\dagger_{le}\hat {\tilde c}_{l+kg},
\end{align}
where $\tilde J=4J\abs{\sin(\phi/2)}/\Delta$. As a result, $\hat H_\text{lab}/\hbar$ becomes:
\begin{align}
    &\hat H_\text{lab}/\hbar\nonumber\\
    &=\frac{\Omega^c}{2}e^{-\im\omega_0t}\sum_{lk\delta}I_{\delta}e^{\im l\phi}e^{-\im(\delta-k)(\pi+\phi)/2}\mathcal J_{\delta-k}(\tilde J)\hat {\tilde c}^\dagger_{le}\hat {\tilde c}_{l+kg}+\text{h.c.}\nonumber\\
    &+\frac{\Omega^s}{2}e^{-\im\omega_1t}\sum_{lk\delta}I_{\delta}e^{\im l\phi}e^{-\im(\delta-k)(\pi+\phi)/2}\mathcal J_{\delta-k}(\tilde J)\hat {\tilde c}^\dagger_{le}\hat {\tilde c}_{l+kg}+\text{h.c.}\nonumber\\
    &+\omega_a\sum_l\hat {\tilde c}^\dagger_{le}\hat {\tilde c}_{le}+\Delta\sum_{l\alpha}\hat {\tilde c}^\dagger_{l\alpha}\hat {\tilde c}_{l\alpha}.
\end{align}

Assuming that the lattice is deep enough that the Wannier states are mostly localized at a single lattice site, and that the carrier and sideband Rabi frequencies are weak and thus cannot drive undesirable transitions, we can set $\delta=0$ and $k=0$, and ignore other terms (see \apdx{acstark} when the $k=0$ approximation is not assumed).  In this case we obtain
\begin{align}
    &H_\text{lab}/\hbar\nonumber\\
    &\simeq\frac{\Omega^cI_{0}}{2}e^{-\im\omega_0t}\sum_{l}\Big(e^{\im l\phi}\mathcal J_0(\tilde J)\hat{\tilde c}^\dagger_{le}\hat{\tilde c}_{lg}\nonumber\\
    &\qquad\qquad\qquad+e^{\im l\phi}e^{\im(\pi+\phi)/2}\mathcal J_{-1}(\tilde J)\hat{\tilde c}^\dagger_{le}\hat{\tilde c}_{l+1g}\Big)\nonumber\\
    &\quad+\frac{\Omega^sI_{0}}{2}e^{-\im\omega_1t}\sum_{l}\Big(e^{\im l\phi}\mathcal J_0(\tilde J)\hat{\tilde c}^\dagger_{le}\hat{\tilde c}_{lg}\nonumber\\
    &\qquad\qquad\qquad+e^{\im l\phi}e^{\im(\pi+\phi)/2}\mathcal J_{-1}(\tilde J)\hat{\tilde c}^\dagger_{le}\hat{\tilde c}_{l+1g}\Big)\nonumber\\
    &\quad+\omega_a\sum_l\hat{\tilde c}^\dagger_{le}\hat{\tilde c}_{le}+\Delta\sum_{l\alpha}l\hat{\tilde c}^\dagger_{l\alpha}\hat{\tilde c}_{l\alpha}.
    \label{eq:Hlabfull}
\end{align}

\section{AC Stark shift in Wannier-Stark OLCs}\label{acstark}
The full Hamiltonian in the lab frame, when driving one of the sideband transitions, reads
\begin{align}
    \hat H^s_\text{lab}/\hbar&=\frac{\Omega^sI_{0}}{2}e^{-\im\omega_1t}\sum_{l}\Big(\mathcal J_0(\tilde J)\hat{\tilde c}^\dagger_{le}\hat{\tilde c}_{lg}\nonumber\\
    &+e^{\im(\pi+\phi)/2}\mathcal J_{-1}(\tilde J)\hat{\tilde c}^\dagger_{le}\hat{\tilde c}_{l+1g}\Big)+\text{h.c.}.
\end{align}
\Rev{Go to the rotating frame as mentioned in the main text, namely, 
\begin{align}
    \hat{\tilde c}^\dagger_{le}\to e^{\im t\left[(\omega_0-\omega_1)l+\omega_0\right]}\hat{\tilde c}^\dagger_{le},~\hat{\tilde c}^\dagger_{lg}\to e^{\im t(\omega_0-\omega_1)l}\hat{\tilde c}^\dagger_{lg},
\end{align}}
when $\omega_0-\omega_1=\Delta$, we get
\begin{align}
    \hat H^s_\text{RF}/\hbar&= \frac{\Omega^sI_{0}}{2}\sum_{l}\Big(e^{\im\Delta t}\mathcal J_0(\tilde J)\hat{\tilde c}^\dagger_{le}\hat{\tilde c}_{lg}\nonumber\\
    &+e^{\im\pi/2}\mathcal J_{-1}(\tilde J)\hat{\tilde c}^\dagger_{le}\hat{\tilde c}_{l+1g}\Big)+\text{h.c.}.
\end{align}
Then we go to the rotating-gauge frame where all the couplings are real:
\begin{align}
    &\hat H^s_\text{RF}/\hbar\nonumber\\
    &=\frac{\Omega^sI_{0}}{2}\sum_{l}\Big(e^{\im\Delta t}\mathcal J_0(\tilde J)\hat a^\dagger_{le}\hat a_{lg}+\mathcal J_{-1}(\tilde J)\hat a^\dagger_{le}\hat a_{l+1g}\Big)+\text{h.c.}\nonumber\\
    &\simeq\frac{\Omega^sI_{0}\mathcal J_{-1}(\tilde J)}{2}\sum_{l}\left(\hat a^\dagger_{le}\hat a_{l+1g}+\text{h.c.}\right)\nonumber\\
    &\qquad-\frac{\left(\Omega^sI_{0}\mathcal J_0(\tilde J)\right)^2}{4\Delta}\sum_{l}\left(\hat a^\dagger_{lg}\hat a_{lg}-\hat a^\dagger_{le}\hat a_{le}\right).
\end{align}
The last line of the above equation accounts for the AC Stark shift. Similarly, we can repeat the above analysis for the carrier drive and obtain
\begin{align}
    &\hat H^c_\text{SSH}/\hbar\nonumber\\
    &=\frac{\Omega^cI_{0}}{2}\sum_{l}\Big(\mathcal J_0(\tilde J)\hat a^\dagger_{le}\hat a_{lg}+e^{-\im\Delta t}\mathcal J_{-1}(\tilde J)\hat a^\dagger_{le}\hat a_{l+1g}\Big)+\text{h.c.}\nonumber\\
    &\simeq\frac{\Omega^cI_{0}\mathcal J_0(\tilde J)}{2}\sum_{l}\left(\hat a^\dagger_{le}\hat a_{lg}+\text{h.c.}\right)\nonumber\\
    &\qquad-\frac{\left(\Omega^cI_{0}\mathcal J_{-1}(\tilde J)\right)^2}{4\Delta}\sum_{l}\left(\hat a^\dagger_{le}\hat a_{le}-\hat a^\dagger_{lg}\hat a_{lg}\right),
\end{align}

In the the main text, we consider a shallow lattice with lattice depth $5E_r$, which gives us $J_0(\tilde J)/J_{1}(\tilde J)\simeq1.73$.

\Rev{The above equations show that the AC Stark shift effectively adds to the system a detuning. When the laser interrogation time is relatively short, we can ignore the AC Stark shift terms if $$\frac{\Omega^sI_{0}\mathcal J_{-1}(\tilde J)}{2}\ll\frac{\left(\Omega^sI_{0}\mathcal J_0(\tilde J)\right)^2}{4\Delta}$$ and $$\frac{\Omega^cI_{0}\mathcal J_0(\tilde J)}{2}\ll\frac{\left(\Omega^cI_{0}\mathcal J_{-1}(\tilde J)\right)^2}{4\Delta},$$ or if $\Omega^c,~\Omega^s\ll\Delta$. However, when the laser interrogation time is long, the AC Stark shift, especially the one related to the sideband drive, cannot be ignored.}

\section{Analytical derivation of $x(T)/a_L$ in the SSH model}
The SSH model in the quasimomentum space reads:
\begin{align}
    \hat H_\text{SSH}/\hbar = \sum_k\left(\frac{\Omega_A}{2}\hat a^\dagger_{ke}\hat a_{kg}+\frac{\Omega_B}{2}e^{-\im ka_L}\hat a^\dagger_{ke}\hat a_{kg}+\text{h.c.}\right),
\end{align}
with eigenenergies $$E^\pm_k=\pm E_k=\pm\frac12\sqrt{\Omega_A^2+\Omega_B^2+2\Omega_A\Omega_B\cos ka_L}$$ and eigenvectors
\begin{align}
    \ket{k,\pm}=\frac{1}{\sqrt{2}}\left(\pm e^{-\im\phi_k}\hat a^\dagger_{ke}+\hat a^\dagger_{kg}\right)\ket{0},
\end{align}
where $$\tan\phi_k=\frac{\Omega_B\sin ka_L}{\Omega_A+\Omega_B\cos ka_L}.$$

The operator $\hat I_y$ and the initial state in k-space are:
\begin{align}
    \hat I_y=-\im\sum_ke^{-\im ka_L}\hat a^\dagger_{ke}\hat a_{kg}+\text{h.c.}
\end{align}
and
\begin{align}
    \ket{\psi_0}&=\hat a^\dagger_{0g}\ket{0}=\frac{1}{\sqrt{L}}\sum_k\hat a^\dagger_{kg}\ket{0}\nonumber\\
    &=\frac{1}{\sqrt{2L}}\sum_k\left(\ket{k,+}-\ket{k,-}\right).
\end{align}
We thus obtain 
\begin{align}
    I_y(t)&=-\im\braket{0|\hat a_{kg}e^{\im Ht}\sum_pe^{-\im pa_L}\hat a^\dagger_{pe}\hat a_{pg}e^{-\im Ht}\hat a^\dagger_{kg}|0}+\text{c.c}\nonumber\\
    &=\frac{1}{2L}\sum_k\left(\frac{\Omega_A\cos{ka_L}+\Omega_B}{E_k^2}\right)\sin(2E_kt)
\end{align}
Thus
\begin{align}
    &\frac{x(T)}{\Omega_Ba_L}=\int_0^TdtI_y(t)\nonumber\\
    &=\frac{1}{2L}\sum_k\left(\frac{\Omega_A\cos{ka_L}+\Omega_B}{E_k^2}\right)\left[1-\cos(2E_kT)\right]\\
    &\xrightarrow[]{L\to\infty}\frac{a_L}{8\pi}\int_{-\pi}^{\pi}dk\left(\frac{\Omega_A\cos{ka_L}+\Omega_B}{E_k^2}\right)\nonumber\\
    &\qquad\qquad\qquad\qquad\times\left[1-\cos(2E_kT)\right]
\end{align}

We identify the Berry phase from the above equation:
\begin{align}
    \mathcal{A}(k)\equiv-\frac12\frac{d\phi_k}{dk}=-\frac12\frac{a_L\Omega_B(\Omega_A\cos{ka_L}+\Omega_B)}{4E_k^2},
\end{align}
and recall the relation between Berry phase and the winding number, $\mathcal{W}$:
\begin{align}
    \mathcal{W}=-\frac{1}{\pi}\int_{BZ}\mathcal{A}(k)dk.
\end{align}
Thus the winding number relates to $x(T)$ via
\begin{align}
    \frac{x(T)}{a_L}=\frac{\mathcal{W}_\text{SSH}}{2}-\frac{1}{2}\int_{BZ}\frac{dk}{2\pi}\frac{d\phi_k}{dk}\cos(2E_kT).
\end{align}

\section{$I_x$ in the presence of small $\delta,~\delta_t$}\label{Ixlr}
We can use linear response theory/perturation theory to understand  the  effect of non-zero but small detunnigs, $\delta$ and $\delta_t$. Using $\hat O=\hat O_0+\Delta\hat O$, where $\hat O_0$ is the unperturbed operator, and
\begin{align}
    \Delta\hat O \simeq \im\int_0^tds\left[\Delta\hat H(s),\hat O(t)\right],
\end{align}
where both $\Delta\hat H(s)$ and $\hat O(t)$ are in the Heisenberg picture of $H_\text{SSH}$. We thus obtain for the case when the unperturbed Hamiltonian is the SSH model:
\begin{align}
    &\Delta\hat I_x(t)=\im\int_0^tds\left[\hat U^\dagger(s)\Delta\hat U(s),\hat I_x(t)\right],
\end{align}
where $\hat U(s) = e^{-\im\hat H_\text{SSH}s}$. When $\Delta\hat H=-\delta\sum_l\hat S^z_l$,
\begin{align}
    &\Delta\hat I_x(t)=-\im\delta\int_0^tds\left[\hat U^\dagger(s)\hat\sum_l\hat S^z_l\hat U(s),\hat U^\dagger(t)\hat I_x\hat U^\dagger(t)\right]\nonumber\\
    &=\frac{\delta}{2L}\sum_k\frac{\cos(2E_kt)-1}{2E_k}\left[e^{\im(-ka_L+\phi_k)}+\text{c.c.}\right]\hat\sigma^z_k.
\end{align}

\begin{figure}
  \centering
    \includegraphics[width=0.9\linewidth]{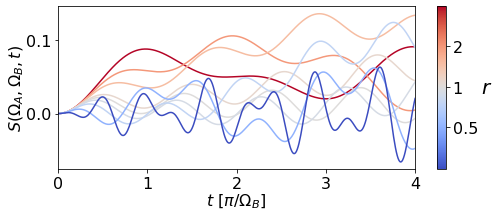}
    \setlength{\abovecaptionskip}{-5pt}
    \setlength{\belowcaptionskip}{-5pt}
    \caption{Numerical results for $S(\Omega_A,\Omega_B,t)$. Here we fix $\Omega_B/(2\pi)=10$ Hz and vary $r$ values.}
   \label{fig:sfcn}
\end{figure}

One can evaluate $\Delta\hat I_x(t)$ at the initial condition $\ket{0}=\ket{0,g}$, and obtain $\braket{\hat I_x(t)}\simeq\delta\mathcal{W}_\text{SSH}/\Omega_B$. Similarly, when $\Delta\hat H=\delta_t\sum_{l,\alpha=e/g}l\hat n_{l\alpha}$, we also have $\braket{\Delta\hat I_x(t)}\propto\delta_t$. In this case,
\begin{align}
    &\int_0^tds\left[\hat U^\dagger(s)\sum_{l,\alpha=e/g}l\hat n_{l\alpha}\hat U(s),\hat U^\dagger(t)\hat I_x\hat U^\dagger(t)\right]\nonumber\\
    &\equiv S(\Omega_A,\Omega_B,t)\equiv\tilde S(\Omega_A,\Omega_B)+\text{osc. term},
\end{align}where $S(\Omega_A,\Omega_B,t),~\tilde S(\Omega_A,\Omega_B)\neq0$ but does not have a direct relation with $\mathcal{W}_\text{SSH}$. We plot numerically evaluated result for $S(\Omega_A,\Omega_B,t)$ in \fig{sfcn}.

\section{Alternative protocol for the SSH clock spectroscopy}\label{sshspec2}

\begin{figure}[h]
    \centering
    \includegraphics[width=0.8\linewidth]{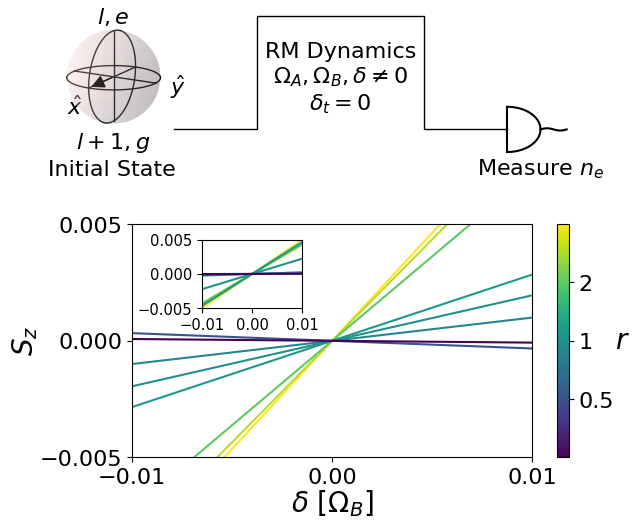}
    \caption{Alternative SSH clock spectroscopy. Upper panel: the protocol. Lower panel: numerical simulation. }
   \label{fig:sshspec2}
\end{figure}

We show in \fig{sshspec2} an alternative protocol for the SSH clock spectroscopy. In this protocol, we first adiabatically
prepare an initial condition \Rev{$(\ket{0,g}+\ket{-1,e})/\sqrt{2}$}, then
turn on the RM dynamics for time $t_B=\pi/\Omega_B$ , followed by
measuring $-S_z$. In fact, with this protocol, we can analytically obtain $\braket{\hat S_z(t)}=-\braket{\hat I_x(t)}$, where $\braket{\hat I_x}$ is precisely given by the results in \apdx{Ixlr}. 

We would like to note that, a combination of adiabatic state preparation the initial state and $S_z$ measurement at the end of the protocol could in principle reduce the statistical noise caused by experimental imperfections.

\Rev{\section{Topological phase transition in finite-size chains }\label{tptLs}

In this section, we briefly study the finite size effects on the topological phase transition. As discussed in the the main text and prior appendices, the sideband coherence of a RM chain relates to the winding number of its underlying SSH model via: $I_x\Omega_B/\delta=\mathcal{W}_\text{SSH}$. This conclusion, however, assumes an infinite SSH/RM chain with periodic boundary condition. Here we discuss the breakdown of this assumption when in reality we have the open boundary condition and finite-size chains. We also discuss how $I_x\Omega_B/\delta=\mathcal{W}_\text{SSH}$ behaves with and without shot-to-shot uncertainty of $\Omega_A,~\Omega_B$, via numerical calculations of  $I_x(t)$ at $t\Omega_B=15\pi$, with fixed $\Omega_B$ and $\delta=0.01\Omega_B$ values. 

We show in \fig{IxLs}-(a)(b) how $I_x\Omega_B/\delta$ varies for  different total number of lattice sites $L$, with initial state being a single atom at $g$ internal state at the center site of the chain. When $L$ is small, the range of $\abs{\Omega_A-\Omega_B}$ values, within which $I_x\Omega_B/\delta$ changes from $1$ to $0$, is very large. As $L$ increases, this range reduces. In our simulation, we observe that for $L>45$, the transition of $I_x\Omega_B/\delta$ becomes very sharp. The energy spectrum of the underlying SSH model is always gapped when $\Omega_A<\Omega_B$ and when $\Omega_A>\Omega_B$.
The energy gap only closes when $\Omega_A=\Omega_B$ as $L\to \infty$. The distinct behavior of $I_x\Omega_B/\delta$ in the regimes $\Omega_A<\Omega_B$ and $\Omega_A>\Omega_B$ is not a consequence of the gap but instead a result of a topological phase transition: such transition happens only in the presence of gap-closing (when $L\to\infty$), and the symmetry of the system remains unchanged before and after the phase transition.

\begin{figure}
    \centering
    \includegraphics[width=\linewidth]{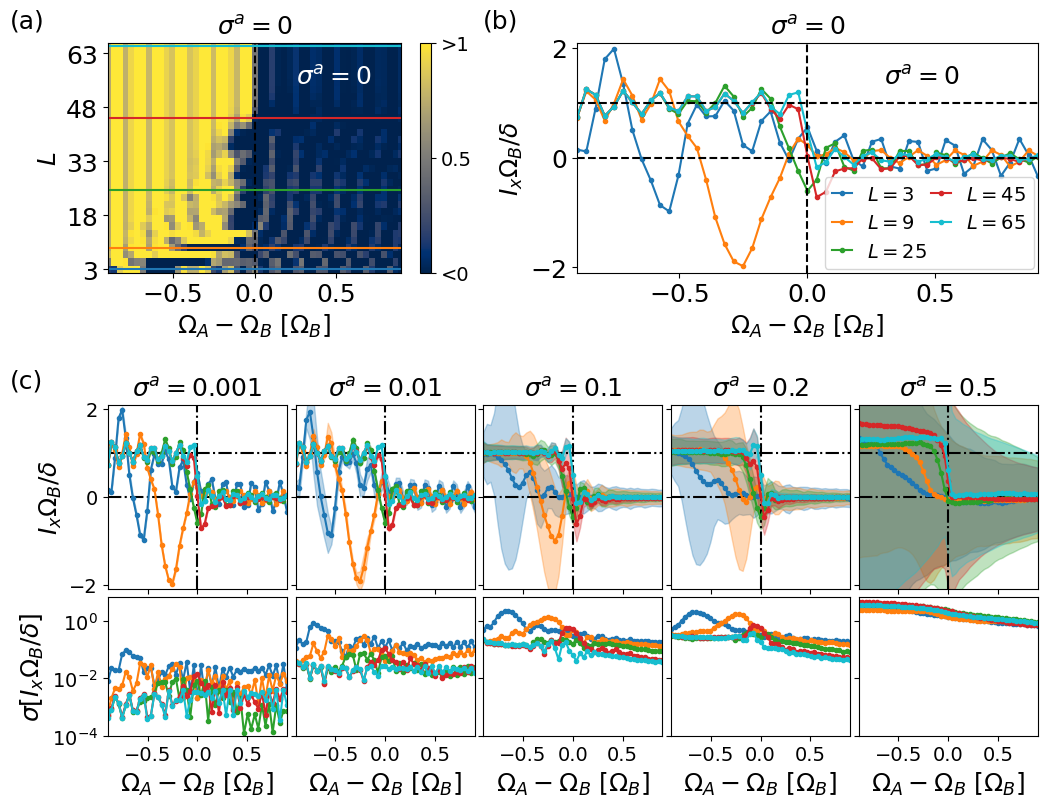}
    \caption{\Rev{Topological phase transition, as indicated by $I_x\Omega_B/\delta$, in a finite chain with length $L$. We vary $\Omega_A$ values, fix $t\Omega_B=15\pi$, $\Omega_B=10$ Hz and $\delta=0.01\Omega_B$. (a)(b): noise-free results of $I_x\Omega_B/\delta$ with different $\Omega_A-\Omega_B$ and $L$ values. (a): $I_x\Omega_B/\delta$ vs $L$ and $\Omega_A-\Omega_B$. (b): several cross-sections of panel (a), as indicated by colors. (c): $I_x\Omega_B/\delta$ in the presence of shot-to-shot amplitude noise of $\Omega_{A/B}$ with standard deviation $\sigma_a\Omega_{A/B}$, where $\sigma_a=0.001,~0.01,~0.1,~0.2,$ and $0.5$. Upper panels: same as (b), but over $500$ realizations of noise. Solid curves showing the average value while shaded areas indicates the standard deviation. Lower panels: values of the standard deviation of the upper panels for better visualization.}}
    \label{fig:IxLs}
\end{figure}

In the presence of  shot-to-shot amplitude variations of the  Rabi frequency. $I_x\Omega_B/\delta$ behaves qualitatively similar to the noise-free case  \fig{IxLs}-(b), but as shown in \fig{IxLs}-(c),  the  standard deviation in   $\sigma[I_x\Omega_B/\delta]$ increases as indicated by the shaded areas in the upper panels and their corresponding lower panels. While  $\sigma[I_x\Omega_B/\delta]$ increases with increasing noise variance $\sigma_a$, as expected, its fluctuation reduces as $L$ increases. Such behavior persists until $\sigma_a$ is as large as $0.5$, when the noise is so large that breaks such protection.

In summary, we have shown in this appendix that, (1) as $L$ increases, the window of $\abs{\Omega_A-\Omega_B}$ values within which $I_x\Omega_B/\delta$ changes between $0$ and $1$ reduces, i.e. as $L$ increases, the topological phase transition  becomes sharper; (2) at a fixed time, as $L$ becomes large enough (in the case here, $L>45$), a topological phase transition can be observed and when deep in the topological phase, $I_x\Omega_B/\delta$ is less impacted by small uncertainties of $\Omega_A,~\Omega_B$, until such uncertainties become too large to be considered as a perturbation.}

\Rev{\section{Full sensitivity comparison between two clock spectroscopy protocols}\label{specsens}

In this section, we compare the sensitivity of our SSH spectroscopy protocol to that of the Rabi spectroscopy protocol. Since the laser interrogation of our protocol is $5$ times that of the Rabi protocol, we need to compare one measurement in our protocol to $5$ measurements of the Rabi protocol.

In the presence of $N$ non-interacting atoms and a global noise with standard deviation $\sigma$, the sensitivity of measuring an observable $O=\sum_iO_i$ where $O_i$ is a single particle observable, for $m$ times, is given by
\begin{align}
    \Delta^2\delta_1=\frac{mN/4+m\gamma^2_1\sigma^2N^2}{m^2(dO/d\delta)^2}=\frac{1}{4mN\alpha^2_1}+\frac{\gamma^2_1\sigma^2}{m\alpha^2_1}, 
\end{align}
where $dO/d\delta=N\alpha_1$, and $\gamma_1$ is how $O$ varies in the presence of noise with standard deviation $\sigma$, or $\partial O/\partial\Omega$ in our case.

When $m=1$, the above expression becomes
\begin{align}
    \Delta^2\delta_2=\frac{N/4+\gamma^2_2\sigma^2N^2}{(dO'/d\delta)^2}=\frac{1}{4N\alpha^2_2}+\frac{\gamma^2_2\sigma^2}{\alpha^2_2}, 
\end{align}
where $dO'/d\delta=N\alpha_2$. For Protocol 2 to have better sensitivity than Protocol 1 when $\Delta^2\delta_1>\Delta^2\delta_2$, we need 
\begin{align}
    \left[\left(\frac{\gamma_1}{\gamma_2}\right)^2\frac{\beta^2}{m}-1\right] N > \frac{1}{4\sigma^2\gamma_2^2}\left(1-\frac{\beta^2}{m}\right)
\end{align}
where $\beta=\alpha_2/\alpha_1$. When we have two protocols with comparable signal and when $\gamma_1\gg\gamma_2$, the above equation becomes
\begin{align}
    N > \frac{m-\beta^2}{4\sigma^2(\gamma^2_1\beta^2-m\gamma^2_2)}
\end{align}

When Protocol 1 is the Rabi spectroscopy and Protocol 2 is the SSH spectroscopy, we have $\sigma=0.001$, $\beta\simeq0.6$, $m=5$, $\gamma^2_2\simeq0.00027$, $\gamma^2_1\simeq0.035$, resulting in $N>10^8$ atoms. Namely, in the presence of over $10^8$ atoms, the SSH protocol has better sensitivity.}

\Rev{We would like to additionally comment that, if both protocols have the same $dO/d\delta$, i.e., $\beta=1$, while all other quantities remain unchanged, we have $N>3\times10^7$. If $m=1$, while all other quantities remain unchanged, we have $N>1.5\times10^7$.}
\bibliography{cmp}

\end{document}